\documentclass[%
 reprint,
 amsmath,amssymb,
 aps,
]{revtex4-2}

\usepackage{graphicx}
\usepackage{dcolumn}
\usepackage{bm}

\usepackage{xcolor}

\begin{document}

\preprint{APS/123-QED}

\title{Hexagonalization in $AdS_3 \times S^3 \times T^4$: Mirror Corrections}

\author{Matheus Fabri}
\email{m.fabri@unesp.br}
\affiliation{%
 Instituto de Física Teórica - UNESP,
 Rua Dr Bento Teobaldo Ferraz 271, 01140-070, São Paulo, Brazil
}%


\date{September 5, 2022}

\begin{abstract}
A big open problem in $AdS_3 \times S^3 \times T^4$ holographic duality is to compute the CFT data of the dual theory. In this direction in \cite{Eden:2021xhe} it was introduced the hexagonalization framework in the $AdS_3$ context. It allows the computation of the structure constants of the CFT$_2$ dual in the planar limit non-perturbatively, however in \cite{Eden:2021xhe} it was introduced only the asymptotic part of the hexagon valid for correlators with asymptotically large bridge lengths. In this work we complete this picture by computing the so called mirror corrections that allow to describe structure constants for finite bridge lengths and as a byproduct we also prove that the half-BPS operators in the theory do not receive these corrections. We end up by giving the first steps on using hexagonalization to compute $n$-point functions in the $AdS_3 \times S^3 \times T^4$ holographic duality.
\end{abstract}

\maketitle


\section{Introduction}

The use of quantum integrability for the $AdS_5 \times S^5$ holographic duality spans over a decade of success now. It lead to the establishment of the gauge-gravity duality as a concrete tool to the analysis of strongly coupled quantum field theories. It allowed the computation of the spectrum of single trace operators in $\mathcal{N}=4$ super Yang-Mills and later to the analysis at finite coupling of the structure constants and $n$-point correlation functions in this theory \cite{Beisert:2010jr,Gromov:2013pga,Basso:2015zoa,Eden:2016xvg,Fleury:2016ykk}. One of the possible paths to analyze now would be how to make use of these techniques in less supersymmetric backgrounds like for example the cases of $AdS_4/$CFT$_3$ and $AdS_3/$CFT$_2$ dualities.

In this work we will analyze the case of $AdS_3/$CFT$_2$ or more specifically the $AdS_3 \times S^3 \times T^4$ background. This background is integrable and the  spectral problem has been analyzed in depth \cite{Sfondrini:2014via,Lloyd:2014bsa,Baggio:2017kza,Borsato:2016xns,Cagnazzo:2012se,Dei:2018mfl,Frolov:2021bwp}. There are some particular points that makes $AdS_3/$CFT$_2$ special and worth analyzing. First of all, the string theory in this background allows the presence of NSNS and RR flux which yields two parameters in the integrable theory that are, respectively, $\kappa$ and $h$ \cite{Lloyd:2014bsa}. For the pure RR case ($\kappa = 0 $) the integrability remains similar to the $AdS_5 \times S^5$ scenario. However for the pure NSNS case ($h=0$) we have a generalization of the flat space superstrings' S-matrix. In this exceptional case the string theory is \textit{exactly} given by a quantum integrable spin chain and the TBA is solved in closed form \cite{Dei:2018mfl} and the model can be also solved by worldsheet CFT techniques \cite{Maldacena:2000hw}. 

In this work we focus on the \textit{pure RR} case that although similar to  $AdS_5 \times S^5$ integrability it has important differences to it. Firstly we have four supermultiplets in this model. There are two massive and two massless multiplets. The latter ones are the main distinction to $AdS_5 \times S^5$ integrability. Their existence makes the comparison with the string theory a little tricky and introduces novel features to the integrable structure. For example, usually the algebraic Bethe ansatz (ABA) yields the spectrum of asymptotically large operators. Finite length corrections are given by the so called wrapping corrections that are suppressed for large operators. However here this is not exactly true because massless modes' wrapping corrections to the spectrum enters at the same order of the ABA \cite{Abbott:2015pps}. Nonetheless quantum integrability can be applied to analyze the spectral problem in this background and more recently even a QSC formalism, to compute the spectrum of finite length operators, was developed \cite{Cavaglia:2021eqr,Ekhammar:2021pys}. However here the story still not complete since the massless modes still elusive in this formalism. For pure RR or NSNS flux the spectral problem is well developed, however for the mixed flux it remains an open problem due to the fact that the dressing phases for the S-matrices are unknown in this case \cite{Frolov:2021fmj,Hoare:2013pma, Lloyd:2014bsa}. 

Given the analysis of the spectral problem in terms of the ABA one can wonder what is the dual CFT$_2$ whose spectrum it describes. From symmetry arguments what is known is that the theory has $\mathcal{N}=(4,4)$ supersymmetry and $\mathfrak{su}(2)_{L} \oplus \mathfrak{su}(2)_{R}$ R-charge. A big open question in $AdS_3 \times S^3 \times T^4$ duality is what is the dual CFT$_2$ for generic mixed flux. Since it is not known for these cases any computation of its CFT data would be highly desirable since it would allow access to this unknown theory. Not everything is lost since it was proved that for unit NSNS flux ($\kappa = 1$ and $h=0$) the dual CFT$_2$ is exactly the symmetric product orbifold of $T^4$ \cite{Eberhardt:2018ouy}. Then in general it is expected that the dual CFT$_2$ is a deformation of this model. For pure NSNS theories and $\kappa\geq 2$ some proposals are given in \cite{Eberhardt:2019qcl,Eberhardt:2021vsx}, however the subject stills open to debate. For mixed flux and pure RR flux points in the moduli space the honest answer is that we do not know what is the dual CFT$_2$. Some proposals for the pure RR case are \cite{Pakman:2009mi,OhlssonSax:2014jtq}. 

To attack the problem of computing structure constants using integrability we have the hexagonalization formalism. In \cite{Eden:2021xhe} the authors introduced hexagonalization in $AdS_3 \times S^3 \times T^4$ background for the mixed flux case. However it was introduced only the asymptotic part of the hexagon which computes structure constants for very large bridge lengths. Thus it remained open the problem of corrections for finite length bridges. These are essential to complete the framework and to compute $n$-point correlation functions. These corrections are the so called mirror corrections.

In this work we attack the problem of defining mirror corrections in $AdS_3 \times S^3 \times T^4$ hexagonalization, thus completing the scheme introduced in \cite{Eden:2021xhe}. More precisely we introduce the mirror corrections and analyze their coupling dependence. We also prove that all the half-BPS states (chiral ring) do not receive wrapping corrections which is a non-trivial fact in this model due to the structure of the chiral ring. Also we provide a first step on computing $n$-point correlation functions for the pure RR CFT$_2$ dual in $AdS_3 \times S^3 \times T^4$ holographic duality and discuss problems one may face when computing the four point function by hexagonalization in this background. 

This paper is organized as follows. In Section \ref{sec:review} we review the integrability setup in $AdS_3 \times S^3 \times T^4$. Section \ref{sec:mirror corrections} we construct the mirror corrections and establish their coupling dependence. In Section \ref{sec:chiral ring} we prove that the chiral ring in this theory does not receive mirror corrections and in Section \ref{sec:correlators} we make some comments on how to compute four point functions in this background. We end up in Section \ref{sec:conclusion} by pointing out possible future directions that would be worth analyzing. For the appendices we leave some technical details.

\section{Integrability in $AdS_3 \times S^3 \times T^4$}
\label{sec:review}

The hexagonalization in $AdS_3 \times S^3 \times T^4$ works in remarkably similar way to its higher dimensional pair. The idea consists in interpreting the structure constants at the planar level as a worldsheet with the topology of a pair of pants and then split it into two hexagon form factors. The operators are given by excited states in the associated spin chain. Then the excitations are distributed in many ways in the two hexagon form factors when the cutting is made and thus we have to sum over all partitions of magnons into two sets for each operator. 

Let's describe the integrability setup in $AdS_3 \times S^3 \times T^4$ more concretely. For this background the isometries are $\mathfrak{psu}(1,1|2)^{\oplus 2}$ and after lightcone gauge fixing the symmetry algebra is given by $\mathfrak{psu}(1|1)_{c.e.}^{\oplus 4}$ centrally extended. This is supplemented by $\mathfrak{u}(1)^{\oplus 4}$ from the $T^4$. Therefore the spectral problem is solved by a $\mathfrak{psu}(1,1|2)^{\oplus 2}$ spin chain with the magnon's S-matrix fixed by the reduced algebra $\mathfrak{psu}(1|1)_{c.e.}^{\oplus 4}$ such that as usual totally fixes the S-matrix for theories with mixed and pure RR fluxes \footnote{The S-matrix for theories with pure NSNS flux is known for an arbitrary $\kappa$ integer. However it is not fixed by symmetry due to the vanishing of central charge that is proportional to $h$ \cite{Dei:2018mfl}.}. The full S-matrix for mixed flux is given in \cite{Eden:2021xhe}. Here we will choose states without momentum and winding in the four torus, therefore $\mathfrak{u}(1)^{\oplus 4}$ is enhanced to $\mathfrak{su}(2)_\bullet \oplus \mathfrak{su}(2)_\circ$. The supercharges carry an index $\alpha=1,2$ of $\mathfrak{su}(2)_\bullet$ and the massless excitations come in two multiplets with each one having an index $\Dot{\alpha}=1,2$ of $\mathfrak{su}(2)_\circ$. We will work in the pure RR limit which consists in taking $\kappa = 0 $ and arbitrary $h$ in the expressions of \cite{Eden:2021xhe}. The main reason is that in this limit is where we have a hexagonalization proposal and more control over integrability. For example for the pure RR regime there is proposals in \cite{Frolov:2021fmj} for the S-matrix dressing factors that are necessary to define the hexagonalization scalar factors at finite coupling. 

\begin{figure}
\includegraphics[width=0.45\textwidth]{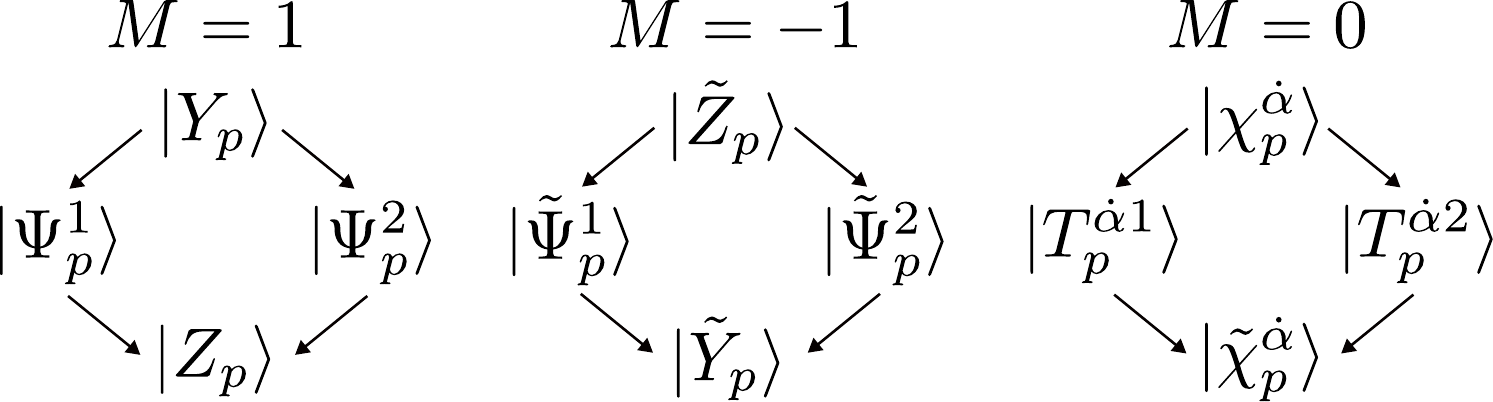}
\caption{Multiplets of fundamental excitations in pure RR $AdS_3 \times S^3 \times T^4$. Here for simplicity we omitted the supercharges that link the states, these can be seen in \cite{Eden:2021xhe}. Note that the bound states also split into left ($M\geq 2$) and right multiplets ($M\leq -2$) that in organize in the exact same way as the fundamental ones.}
\label{fig:multiplets}
\end{figure}

The excitations carry a $\mathfrak{u}(1)$ charge $M$ that is quantized in the pure RR model and classifies the four  supermultiplets. For $M=1$ we have the left (L) multiplet with $Y$ being the $S^3$ mode, $Z$ the $AdS_3$ mode, and the superpartners $\Psi^{\alpha}$. The other massive multiplet with $M=-1$ is the right (R) multiplet with the respective excitations $\tilde{Y}$, $\tilde{Z}$, and $\tilde{\Psi}^{\alpha}$. These L and R labels comes from the fact that the charge $M$ is given by
\begin{equation}
    M = L_{0} - \tilde{L}_{0} - J_3 + \tilde{J}_3,
\end{equation}
where $\{L_0, \tilde{L}_0\}$ are the left and right dilatation operators of the conformal group and $\{J_3, \tilde{J}_3\}$ the related R-charge generators. Then the sign of $M$ would be linked to the chirality in the dual CFT$_2$. The remaining excitations are two multiplets of massless modes ($M=0$) consisting of fermions $\chi^{\dot{\alpha}}$, the $T^4$ bosons $T^{\dot{\alpha}\alpha}$, and the fermions $\tilde{\chi}^{\dot{\alpha}}$. For $|M|\geq 2$ the states are then bound states of these $|M|=1$ fundamental magnons. The spectrum of fundamental particles in showed in Fig. \ref{fig:multiplets}. As usual these excitations can be split into a tensor product of states in $\mathfrak{psu}(1|1)_{c.e.}^{\oplus 2}$ as detailed in \cite{Eden:2021xhe}.

For the massive modes integrability in the pure RR limit works out in a very similar way to $AdS_5 \times S^5$. However for the massless modes there are some distinctions. The first one is the kinematics of the mirror theory. Consider the dispersion relation of a single magnon with momentum $p$ in the pure RR limit 
\begin{equation}
    E(p,M) = \sqrt{M^2 + 4h^2 \sin(p/2)^2}.
\end{equation}
As usual we can introduce the Zhukovsky variables $x(u)$ in terms of a spectral parameter and analyze the kinematics with respect to it. Then the energy and momentum become rational functions of the Zhukovsky variables. For massive particles a mirror transformation is an analytic continuation in the spectral parameter $u$ which has the effect of transforming the Zhukovsky variables. But for massless modes it appears that no simple transformation on the Zhukovsky variables is possible. The kinematics of the mirror theory is central to the definition of the mirror correction in the next section, therefore in the App. \ref{app:mirror} we detailed its construction for massive and massless modes. 

Another difference of this background is the set of half-BPS states. It is clear that for the massless modes at zero momentum there is no cost in energy for adding them. This is not new, usually zero momentum magnons in the spin chain corresponds to a descendant state after a symmetry transformation.  This is not the case when we add massless fermions at $p=0$. These are not descendants and their insertions yield 16 possible families of half-BPS operators that are denominated the \textit{chiral ring} \cite{Baggio:2017kza}. This is in sharp contrast to  $AdS_5 \times S^5$ where the half-BPS operators are classified only by their orbital R-charge. Nonetheless the scaling dimension and structure constants of the chiral ring remain protected. The chiral ring is detailed in Section \ref{sec:chiral ring}.

After this detour in the integrable setup let's come back to hexagonalization. The rules for gluing the form factors are simple for large bridges connecting the operators. In this case the we just multiply the hexagon form factors for each partition. If the bridges are finite one has to correct the prescription by adding mirror corrections. How to do this in the context of $AdS_3 \times S^3 \times T^4$ is what we discuss in the next section. 

\section{Adding mirror corrections}
\label{sec:mirror corrections}

In \cite{Eden:2021xhe} it was described only the asymptotic part of the $AdS_3 \times S^3 \times T^4$ hexagon. There are two types of wrapping corrections one has to add to have the complete picture. The first is due to finite size bridges and the second is due to finite size operators \cite{Basso:2015zoa,Basso:2017muf,Basso:2022nny}. It is the former that we describe here how to insert in $AdS_3$ hexagon framework. For the latter we note that it certainly appears just like in the $AdS_5$ case as poles that have to be regularize as a mirror magnon circles one excited operator. However it may be more useful to pursuit it after one has a more appropriate control over the TBA for this background, more knowledge about the prescription to compute the dressing phases and control over the dual CFT analysis to have data to compare.

\begin{figure*}
\includegraphics[width=0.7\textwidth]{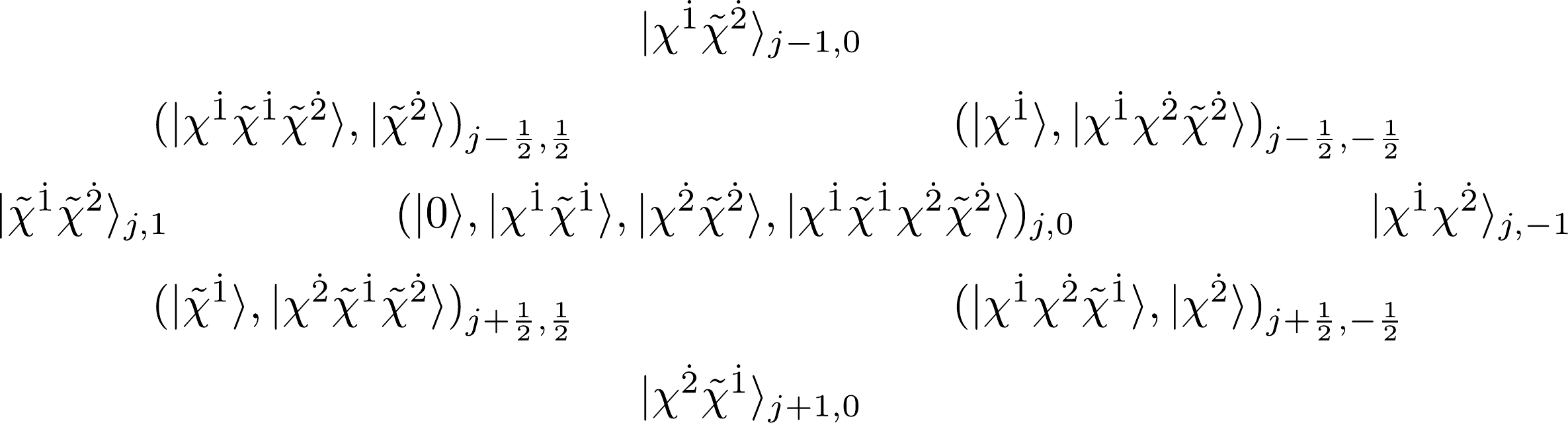}
\caption{All 16 families of half-BPS states created on top of a vacuum of length $j$. Note that all massless fermions are inserted with their respective zero momentum limit and the subscripts denote their scaling dimension and R-charge, respectively.}
\label{fig:bps-states}
\end{figure*}

The wrapping corrections are added in the hexagon picture by introducing a complete pair of states of mirror particles in each cut of the hexagon. Then these states are added to each one of the hexagons and we sum over the flavours and bound state numbers of the inserted particles. Unlike the $AdS_5$ case there are three sets of particles that one must insert, left and right multiplets, and massless particles. Only the former can make bound states with bound state number $Q$ and $\bar{Q}$, respectively. The general form of the massive mirror corrections for a single particle insertion are
\begin{align}
      \sum_{Q=1}^{\infty} \int_{\mathbb{R}} \frac{du}{2\pi} \mu_{Q} (u^\gamma) e^{-\tilde{E}_{Q}(u)l}\ \textrm{int}^{L}_{Q} (u^{\gamma} |\{ u_i \}) + \nonumber \\
     + \sum_{\bar{Q}=1}^{\infty}   \int_{\mathbb{R}} \frac{du}{2\pi} \mu_{\Bar{Q}} (u^\gamma) e^{-\tilde{E}_{\bar{Q}}(u)l}\ \textrm{int}^{R}_{\bar{Q}}(u^{\gamma} |\{ u_i \}),  
\end{align}
and for massless mirror corrections
\begin{equation}
    \int_{|u|\geq 2} \frac{du}{2\pi} \mu_{\circ} (u)  e^{-\tilde{E}_{\circ}(u)l} \ \textrm{int}^{\circ}(u|\{ u_i \}),
\end{equation}
where $l$ is the associated bridge length, $\tilde{E}_{*}$ is the mirror energy and $\mu_{*}(u^{\gamma})$ the mirror measure of the insertion. Note that $u^{\gamma}$ denote a mirror transformation and $u^{2\gamma}$ a crossing transformation following the usual notation as in \cite{Basso:2015zoa}. The mirror dynamics and bound states of the model are described in App. \ref{app:mirror} and App. \ref{app:bs}, respectively. Then there are two ingredients to find: the mirror measure and the integrand.

Using the one particle hexagon normalizations of \cite{Eden:2021xhe} the measure is defined then as
\begin{equation}
    \frac{1}{\mu_A (v)} = \textrm{res}_{u\rightarrow v} \langle \mathfrak{h}| \Bar{A}_{u^{2\gamma}} A_{v} \rangle.
\end{equation}
Where $A$ is any particle flavour and $\Bar{A}$ the crossed particle. There are some differences from the $AdS_5$ measure. First there are four supermultiplets, then in principle one has four measures. But we assume that the matrix elements are blind to $su(2)_\circ$, this makes the measure of massless supermultiplets equal. Also using the L/R-symmetry we equate the left and right associated measures. Thus in the end we have two independent measures to compute \footnote{To calculate these measures we used the $\mathfrak{psu}_{c.e.}(1|1)^{\oplus 2}$ S-matrices defined in \cite{Eden:2021xhe}.}.

For the massive part of the spectrum of fundamental particles the measure is given by \footnote{Note that to compute this measure we used the crossing equations and braiding unitarity. Therefore the measures computed here are independent of the issues raised in \cite{Frolov:2021fmj} since we used only these properties and not the explicitly form of the dressing factors.}
\begin{equation}
    \frac{1}{\mu (v)} = \textrm{res}_{u\rightarrow v} \frac{x^{+}_{u}}{x^{-}_{u}} \langle \mathfrak{h}| \Tilde{Y}_{u^{2\gamma}} Y_{v} \rangle = \frac{x^{+}_{v}}{x^{-}_{v}} \textrm{res}_{u\rightarrow v} A^{RL}(u^{2\gamma} , v).
\end{equation}
Where we added a momentum factor that comes from crossing and is related to the spin chain frame choice \footnote{This is detailed in the App. F of \cite{Basso:2015zoa} and is the same for the $AdS_3$ case.}. From this definition and the mirror kinematics one can derive the measure for the massive modes (in the mirror region):
\begin{align}
    \mu_{a}(u^\gamma) & = \frac{ x^{[+a]}_{u}x^{[-a]}_{u}}{h \left( 1 -x^{[+a]}_{u}  x^{[-a]}_{v} \right)} \nonumber \\
   & \times \sqrt{\left(1-\left(x^{[-a]}_{u}\right)^2\right)\left(1-\left(x^{[+a]}_{u}\right)^2\right)}.
\end{align}
Where $a$ is the bound state number. To go from fundamental particles to bound states one can use fusion as in $AdS_5$ or simply change the shift in the Zhukovsky variables to $\pm a$. This works because the only difference from fundamental to bound states in the $AdS_3$ is the mass since both transform in the same representation. The massless particles' measure is computed similarly. For it we have:
\begin{equation}
    \frac{1}{\mu_{\circ} (v)} = \textrm{res}_{u\rightarrow v}  i \langle \mathfrak{h}| \Tilde{\chi}^{\dot{A}}_{u^{2\gamma}} \chi^{\dot{B}}_{v} \rangle = \textrm{res}_{u\rightarrow v} D^{\circ\circ}(u^{2\gamma} , v).
\end{equation}
Here there are no momentum factor for crossing since the crossed particles are fermionic. Then:
\begin{equation}
    \mu_\circ (u) = -\frac{(x_u)^2 }{ (1-(x_u)^2)^2}.
\end{equation}
Which is significantly distinct from the massive measure. As seen in App. \ref{app:mirror} to go to the mirror region we just change the range of $u$. So the form of $\mu_\circ (u)$ remains the same for the mirror measure.

There are some significant facts that we can highlight from these expressions. First of all $\mu_{a}(u^\gamma)$ can be seen as the square root of the $AdS_5$ measure. Also $\mu_{\circ}(u)$ is the massless limit of $\mu_{a}(u)$ and not of its mirror transformed version (up to the $1/h$ factor). This makes sense since there is no mirror transformation in the Zhukovsky variables for massless particles. At last but not least, following the weak coupling expressions in App. \ref{app:weak}, one has that the mirror corrections for a bridge of length $l$ given by the measure and energy propagating factors ($l$-dependent part of the integrand) are like
\begin{align}
    & \textrm{\textbf{Massive particles:}}\ \mathcal{O}(h^{1+2l}), \nonumber \\
    & \textrm{\textbf{Massless particles:}}\ \mathcal{O}(h^{0}). \nonumber
\end{align}
Which differs significantly from the $AdS_5$ case since there all loop corrections come in even powers of the integrability coupling. In Section \ref{sec:correlators} we explain how our results are compatible with the OPE.

Since the measure has been dealt with we turn now to the integrand. Following \cite{Basso:2015zoa} we can write it as the transfer matrix of $\mathfrak{psu}(1|1)^2$. Consider for example a single mirror particle insertion. After the sum over flavours of the mirror particle we obtain
\begin{equation}
    \textrm{int}^{*}_{Q} (u^{\gamma} | \{ u_{i} \} ) = \mathcal{A}_{\textrm{asymptotic}} \prod_{j} h_{Q,1}^{**_{j}} (u^{\gamma},u_j ) \ \hat{\Delta}^{*}_{Q} (u^{\gamma} | \{ u_{i} \} )
\end{equation}
Where $\mathcal{A}_{\textrm{asymptotic}}$ is the asymptotic piece of the hexagon, $h^{**_{j}}$ is the hexagon scalar factor for the appropriate representation $*$, and $\hat{\Delta}^{*}_{Q}$ is the eigenvalue of the transfer matrix. From this rewriting in terms of transfer matrices, we can check that the wrapping corrections for massless modes \textit{vanish}. The main argument is that the two massless multiplets are equal, however they come with opposite statistics and therefore they cancel. The eigenvalues of the transfer matrices are given in App. \ref{sec:transfermatrix}.

We can also compute the integrand when we insert multiple mirror particle corrections as in \cite{Basso:2015eqa}. Indeed by using braiding unitarity for the S-matrix we note that for each pair of mirror insertions we must consider a factor like $h(u^\gamma , v^\gamma) h(v^\gamma , u^\gamma)$. Combining this and Yang-Baxter equation we get a product of transfer matrices. An example of a mirror integrand with multiple L-particles is
\begin{align}
   \textrm{int}^{L}_{Q_n} (\{ u^{\gamma}_{n} \} | \{ u_{i} \} ) = \mathcal{A}_{\textrm{asymptotic}} \prod_{k} \prod_{j} h_{Q_k ,1}^{L *_{j}} (u_k^{\gamma},u_j ) \nonumber \\
   \prod_{k\neq j} h_{Q_k ,Q_j }^{LL} (u_k^{\gamma} ,u_j^{\gamma} ) h_{Q_j ,Q_k }^{LL} (u_j^{\gamma} ,u_k^{\gamma} )\ \prod_{k} \hat{\Delta}^{L}_{Q_{k}} (u_k^{\gamma} | \{ u_{i} \} ).
\end{align}
The integrand with distinct multiplets and particle content can easily be generalized. A final point that we want to make is that even though the massless corrections vanish for the structure constant they could play a role in the four point function hexagonalization as we will explain in Section \ref{sec:correlators}. 

\section{Wrapping corrections and the chiral ring}
\label{sec:chiral ring}

The simplest structure constants to compute are the ones involving only half-BPS operators. Unlike in the $AdS_5$ case we have a more involved set of half-BPS operators with nontrivial structure constants. In $\mathcal{N}=4$ SYM the highest weight of a half-BPS multiplet is completely specified only by the R-charge, however here we have 16 possible families of half-BPS operators. The reason for this is that amount of preserved supersymmetry is much smaller than in the $AdS_5$ case. As proved in \cite{deBoer:2008ss,Baggio:2012rr} the chiral ring of the $\mathcal{N}=(4,4)$ supersymmetric dual CFT$_2$ has protected structure constants. So they yield a good laboratory to test the hexagon computations. 

Let's describe these operators in more detail. From the point of view of integrability we construct this chiral ring by adding massless fermionic excitations with null momenta in the original half-BPS vacuum $| 0 \rangle$. There are four massless fermions $\chi^{\dot{A}}$ and $\tilde{\chi}^{\dot{A}}$ that we can add and these furnish the previously mentioned 16 families of the chiral ring. The correspondence between the families of half-BPS operators and zero momentum fermion insertions is described in Fig. \ref{fig:bps-states}.

Since the structure constants of the chiral ring are protected one expect that the wrapping corrections to vanish, at least for massive mirror particles. Here we will prove that in fact they \textit{vanish for all mirror insertions}. The proof is similar to the cancellation of wrapping corrections in the TBA of half-BPS states \cite{Baggio:2017kza}.

\begin{figure}
\includegraphics[width=0.45\textwidth]{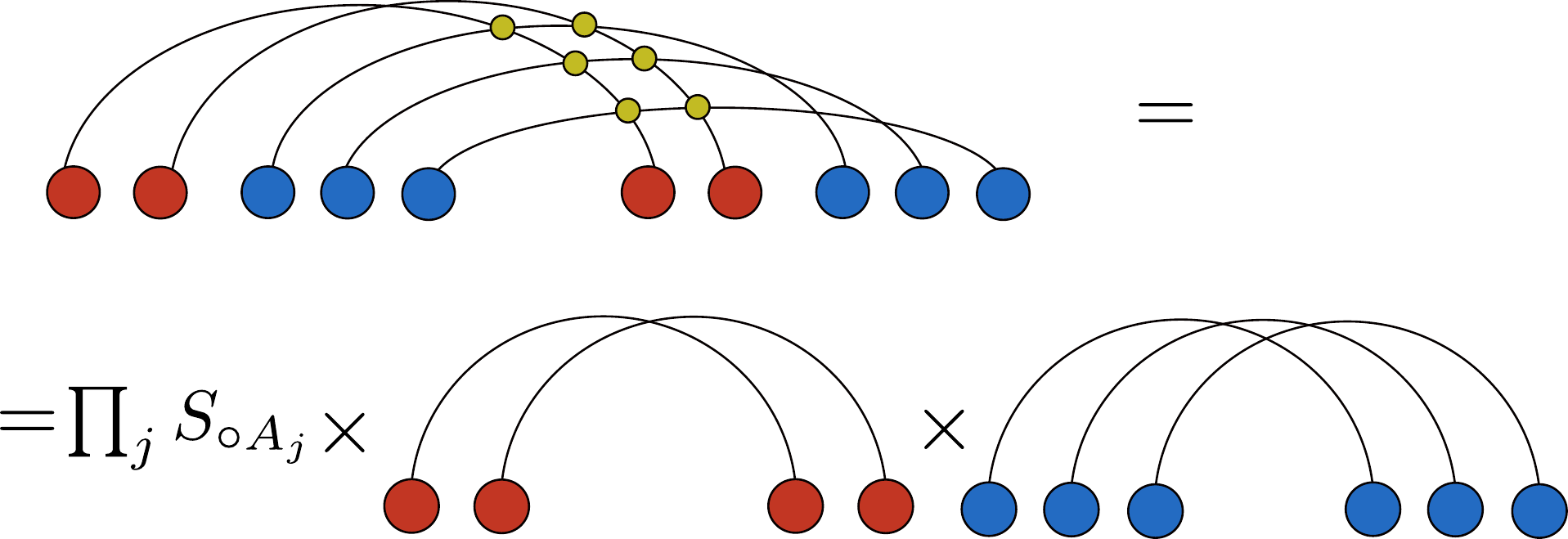}
\caption{Factorization of the hexagon form factor with the insertion of fermionic massless modes at zero momentum. On the left we have an example of a hexagon form factor with massless fermions at null momenta (in red) and some other insertions (in blue). Since these fermions transform as singlets their scattering with the other modes (yellow dots) is pure transmission. Then the hexagon factorizes as a product of phases and the corresponding hexagon form factors.}
\label{fig:massless-hexagon}
\end{figure}

First we need to see the effect of adding these massless fermions at zero momentum. By analyzing the $\mathfrak{psu}(1|1)_{c.e.}^{\oplus 4}$ symmetry one can see that the massless multiplets are annihilated by the algebra for zero momentum particles \cite{Eden:2021xhe}. This is true also for the subalgebra $\mathfrak{psu}(1|1)_{c.e.}^{\oplus 2}$, which is the one that plays a role in the mirror corrections as a transfer matrix. This vanishing implies that null momentum massless particles transform as a singlet and thus scatter purely by transmission with another states. Then the hexagon form factor with these particles have their contribution factorized. Let $X_\circ$ be some fermion at zero momentum. Therefore
\begin{align}
    \langle \mathfrak{h} | X_{\circ} A_1 (p_1) & \cdots A_n (p_n) \rangle = \nonumber \\
   &  \prod_{j} S_{\circ A_j}(p_j) \ \langle \mathfrak{h} | A_1 (p_1) \cdots A_n (p_n) \rangle.
\end{align}
Where $S_{\circ A_j}(p_j)$ is a phase factor that depends on the flavour and momentum of the scattered particle $A_j$ and which massless fermion was added. If there are more than one massless fermion at zero momentum, the hexagon form factor factorizes into a phase too. However now this phase is a product of the phases $S_{\circ A_j}(p_j)$ times the hexagon of only the massless zero momentum fermions insertions as denoted in Fig. \ref{fig:massless-hexagon}. The hexagon with only massless fermions at zero momenta demands a more careful treatment as detailed in \cite{Eden:2021xhe}, but the full hexagon form factor factorizes nonetheless \footnote{This factorization resembles the charged hexagon formulation that appears in the hexagonalization of the fishnet theory in \cite{Basso:2018cvy}.}.

Now that massless fermions can be added in the hexagon in a simple way we can look at the cancellation of wrapping corrections. We note that the phase $S_{\circ A_j}(p_j)$ is actually independent of the flavour of the scattered particles, in fact it depends only on the momentum $p_j$ and the multiplet where the particle lies. The explicit form of these S-matrices is given in App. \ref{sec:phasefactors}. Therefore when a massive mirror magnon is inserted one has that
\begin{equation}
    \hat{\Delta}^{*}_{a} (u^{\gamma} | \{ u_{i} \} ) = \prod_{j} S_{\circ}(p_j)\ \hat{\Delta}^{*}_{a} (u^{\gamma} | \varnothing ) = 0 \ \ \textrm{if} \ \ *=L,R.
\end{equation}
Then all mirror corrections when L and R mirror particles are on top of the half-BPS states vanish. As said before, for massless mirror modes the wrapping corrections always vanish so we disregard them. With this we conclude that all half-BPS three point functions do not receive wrapping corrections and are given only by the asymptotic hexagon prescription.

\section{Speculations on Four point functions}
\label{sec:correlators}

Given now that the mirror corrections to hexagonalization are known one could try to bootstrap the $n$-point functions as it was done for $\mathcal{N}=4$ SYM in \cite{Fleury:2016ykk,Fleury:2017eph}. This would be a very strong result since no alternative exists to compute correlators for the theory dual to the background with pure RR flux. As said before for pure NSNS flux and $\kappa=1$ it is known that the CFT$_2$ dual is the symmetric product orbifold Sym$_N (T^4)$. Correlation functions in this model were computed by using a differential equation method in \cite{Dei:2019iym}. Also diagram techniques to compute these were introduced in \cite{Pakman:2009zz} and these are similar to the large-$N$ t'Hooft expansion in matrix models.

With this success story for pure NSNS backgrounds we will try to bootstrap four point functions for the pure RR model. Before proceeding we should establish some facts of the CFT$_2$ first. The dual of Type IIB syperstrings in $AdS_3 \times S^3 \times T^4$ has $\mathcal{N}=(4,4)$ superconformal symmetry. The operators in this theory are classified by $\mathfrak{su}(2)_L \oplus \mathfrak{su}(2)_R$ R-charge. A general operator is denoted by $\mathcal{O}(x,\Bar{x}|y,\Bar{y})$ where $\{x,\Bar{x}\}$ are their position in holomorphic variables and $\{y,\Bar{y}\}$ their polarizations in $S^3$ \cite{Pakman:2007hn,Rastelli:2019gtj}. We can decompose them in modes with definite $(J^3,\tilde{J}^3)$ magnetic charges:
\begin{align}
    \mathcal{O}(x,\Bar{x}|y,\Bar{y}) = \sum\limits_{m=-J}^{J} \sum\limits_{\Bar{m}=-\Bar{J}}^{\Bar{J}} \sqrt{c_{J,m} c_{\Bar{J},\Bar{m}}}\ y^{J-m} \ \Bar{y}^{\Bar{J}-\Bar{m}} \ \nonumber \\
    \mathcal{O}_{m,\Bar{m}} (x,\Bar{x}) \ \ \textrm{with}  \ \ c_{a,b} = \binom{2a}{a+b}.
\end{align}
Where $(J,\Bar{J})$ are the total left and right R-charges and $(m,\Bar{m})$ are the magnetic charges. Here $y=0$ picks the highest weight state and $y=\infty$ the lowest weight state, for example. To pick descendant states we can apply derivatives in $y$ and $\bar{y}$.

We will be interested in the four point function of half-BPS operators with identical dimensions that are dual to the $|0\rangle$ state in the chiral ring. Since there are no excitations, the correlator will be given purely by the wrapping corrections contribution. We define the conformal and R-charge cross-ratios as
\begin{align}
    \eta = \frac{x_{12} x_{34}}{x_{13} x_{24}} \ \ \ \Bar{\eta} = \frac{\Bar{x}_{12} \Bar{x}_{34}}{\Bar{x}_{13} \Bar{x}_{24}} \label{eq:cross-ratios1} \\
    \alpha = \frac{y_{12} y_{34}}{y_{13} y_{24}} \ \ \ \Bar{\alpha} = \frac{\Bar{y}_{12} \Bar{y}_{34}}{\Bar{y}_{13} \Bar{y}_{24}}.
    \label{eq:cross-ratios2}
\end{align}
Then  this four point function of scalars with dimension $\Delta$ can be written as
\begin{equation}
    G(x_i , y_i) = \left(\frac{y_{12}\bar{y}_{12}}{x_{12}\bar{x}_{12}}\frac{y_{34}\bar{y}_{34}}{x_{34}\bar{x}_{34}}\right) ^{\Delta} \mathcal{G}(\eta,\bar{\eta},\alpha,\bar{\alpha}).
\end{equation}
Where it is the computation of the invariant $\mathcal{G}(\eta,\bar{\eta},\alpha,\bar{\alpha})$ that we are interested in. 

The correlator must satisfy some consistency properties. The first of them is that the correlator must be invariant under $\eta \leftrightarrow \bar{\eta}$ \textit{and} $\alpha \leftrightarrow \bar{\alpha}$ permutations. The fact that we have a CFT$_2$ demands that both exchanges must be made together to have invariance, unlike in higher dimensions. Another requirement is that it must satisfy the superconformal Ward identities \cite{Aprile:2021mvq,Rastelli:2019gtj}:
\begin{align}
    (\partial_{\eta} + \partial_{\alpha})\mathcal{G}(\eta,\bar{\eta},\alpha,\bar{\alpha})|_{\eta = \alpha} = 0, \\ 
    (\partial_{\bar{\eta}} + \partial_{\bar{\alpha}})\mathcal{G}(\eta,\bar{\eta},\alpha,\bar{\alpha})|_{\bar{\eta} = \bar{\alpha}} = 0.
\end{align}
A four point function that satisfy these can be split in the general form
\begin{align}
    \mathcal{G}(\eta,\bar{\eta},\alpha,\bar{\alpha}) = &\  \mathcal{C} + (\eta - \alpha) S(\eta, \alpha) + (\bar{\eta} - \bar{\alpha}) S(\bar{\eta}, \bar{\alpha}) + \nonumber \\
    & + (\eta - \alpha)(\bar{\eta} - \bar{\alpha})\ \mathcal{H} (\eta,\bar{\eta},\alpha,\bar{\alpha}).
\end{align}
Where $\mathcal{C}$ is a constant, $S(\eta, \alpha)$ is related to short multiplets of the superalgebra, and $\mathcal{H} (\eta,\bar{\eta},\alpha,\bar{\alpha})$ is related to long multiplets. Therefore whatever is the correlator one finds it musts satisfy these two requisites for consistency. These are necessary but not sufficient conditions.

\begin{figure}
\includegraphics[width=0.45\textwidth]{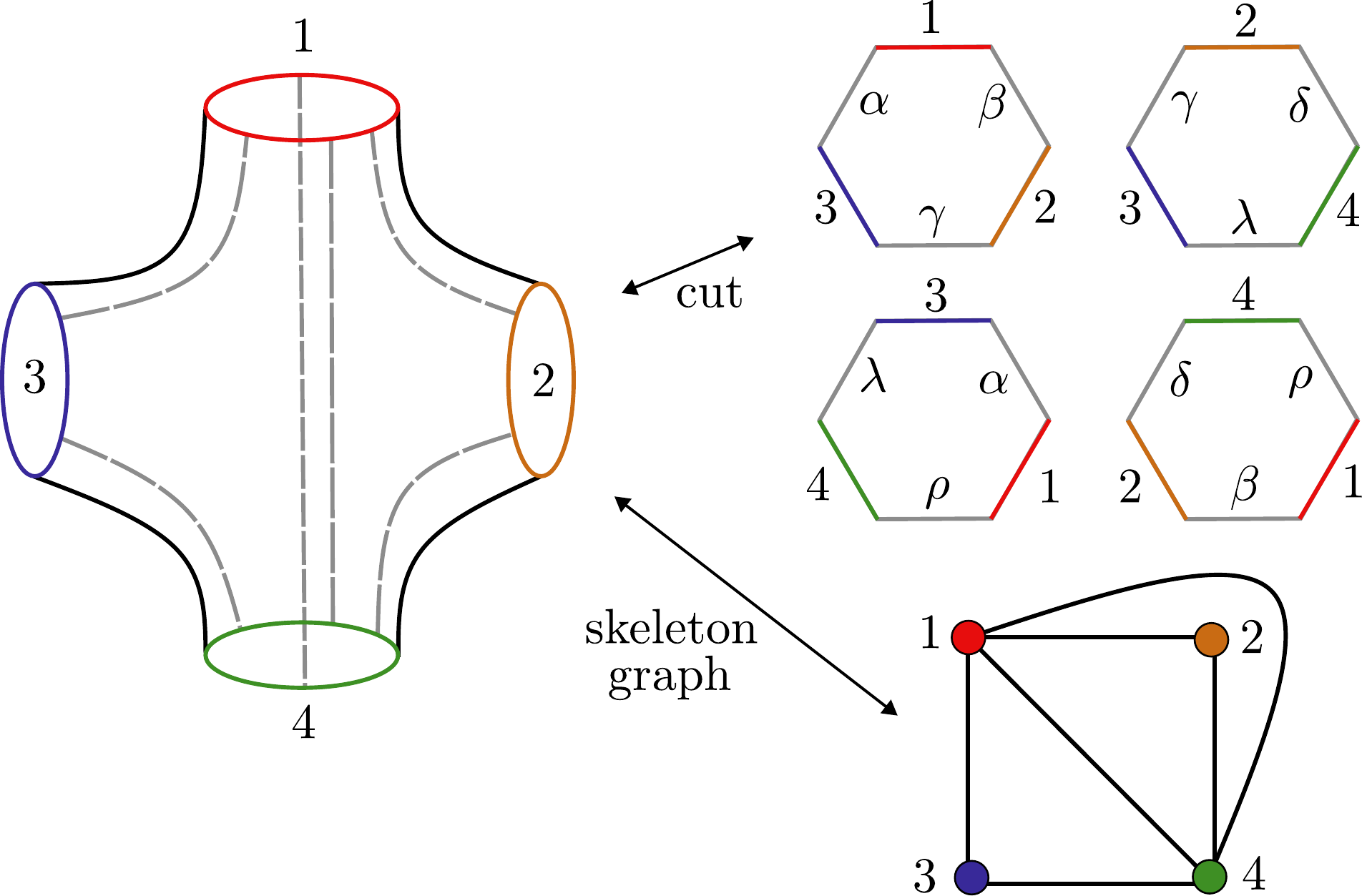}
\caption{Hexagonalization of a four point function and the related skeleton graph and hexagon decomposition. Here we consider a specific hexagonalization cutting of a four point function. On the left we have the four point function where it is the dashed lines that we cut it. In the left we have all the four hexagons that decompose the four point function such that we must identify and then glue all edges that have the same greek letter label. Also related to this specific cutting we have the skeleton graph on the below left.}
\label{fig:four-point}
\end{figure}

The hexagonalization of four point functions works in a similar way to the structure constant one. Basically the worldsheet corresponding to the correlator is cut into four hexagonal patches as described in Fig. \ref{fig:four-point}. The rules for gluing these suffer a small modification from before. Since there are now hexagons with distinct operators in their physical edges we can not put all of them in the canonical frame at the same time. Thus there is a weight $\mathcal{W}_{ij}$ in all bridges $ij$ related to the factor one gains from gluing an hexagon in the canonical frame and another one that is polarized. More details can be seen in \cite{Fleury:2016ykk} in the context of $AdS_5 \times S^5$.

There are many ways to cut the four point correlator into hexagons. The usual prescription is to associate a specific cutting to a skeleton graph and then sum over all skeleton graphs. Where these are the graphs coming from contracting the correlators at weak coupling. Then one could propose as in \cite{Fleury:2016ykk} that the correlator are given by
\begin{align}
    G(x_i,y_i) = \sum_{\Gamma} & \left(\prod_{(ij)} d_{ij}^{L_{ij}} \right) \times \nonumber \\
    & \left[ \sum_{\psi_{(ij)}} \int d\mu_{\psi_{ij}} (u_{ij})\ \mathcal{
    W}_{\psi_{ij}} \prod_{(ijk)} \mathcal{H}_{(ijk)} \right],
\end{align}
where $\Gamma$ is the set of all skeleton graphs,
$\mathcal{H}_{(ijk)}$ is hexagon contribution with mirror particles inserted corresponding to face $(ijk)$, and $d\mu_{\psi_{ij}}$ the associated measure to the mirror particle. An example of a relation between a specific hexagonalization of the four point function and the related skeleton graph is given in Fig. \ref{fig:four-point}. Note that each skeleton comes with a weight that is just the contribution of the skeleton graph in perturbation theory. A conservative proposal for a 2D CFT would be
\begin{equation}
    d_{ij} = \frac{y_{ij}}{x_{ij}}\frac{\bar{y}_{ij}}{\bar{x}_{ij}}.
\end{equation}
Which is just the contraction of scalar fields of R-charges $J_3 = \tilde{J}_3 = 1/2$ and dimension $\Delta=1$. For the moment we assume that this prescription of assigning skeleton graphs is also valid here in $AdS_3 \times S^3 \times T^4$. The validity of this assumption will be discussed in the end in the concluding remarks (Section \ref{sec:conclusion}).

We have that $\mathcal{W}_{\psi}$ is the eigenvalue of the rotation matrix between a canonical and non-canonical hexagon. Following the analysis of \cite{Fleury:2016ykk} we put the hexagon containing operators $\mathcal{O}_1$, $\mathcal{O}_3$ and $\mathcal{O}_4$ at the canonical configuration in Fig. \ref{fig:four-point} which implies that
\begin{equation}
    \eta = x_2 \ \ \ \Bar{\eta} =\Bar{x}_{2} \ \ \ \textrm{and} \ \ \ \alpha = y_2 \ \ \ \Bar{\alpha} =\Bar{y}_{2}.
\end{equation}
Then the change of frame is given by the dilation operators $\{ L_0 , \tilde{L}_0 \}$ for $\{\eta ,\bar{\eta} \}$ and R-charges $\{ J_3 , \tilde{J}_3 \}$ for $\{\alpha ,\bar{\alpha} \}$. The weight is just the holomorphic and antiholomorphic scalings and their sphere counterparts:
\begin{equation}
    \mathcal{W}_{\psi} = e^{-L_{0} \log\eta}e^{-\tilde{L}_{0} \log\bar{\eta}} e^{J_{3} \log\alpha}e^{\tilde{J}_{3} \log\bar{\alpha}}.
\end{equation}
This is the weight we acquire one gluing the edge $1-4$ in Fig \ref{fig:four-point}. For other edges we simply exchange the operator indices in (\ref{eq:cross-ratios1}) and (\ref{eq:cross-ratios2}) to obtain the appropriate weight for that edge. We can rewrite this in terms of more adequate variables that are the energy $E$ and mass $M$ of the magnons
\begin{align}
    E = L_{0} + \tilde{L}_{0} - J_3 - \tilde{J}_3,\\
    M = L_{0} - \tilde{L}_{0} - J_3 + \tilde{J}_3.
\end{align}
So the weight acquires a convenient form for us:
\begin{equation}
    \mathcal{W}_{\psi} = e^{-\frac{E}{2}\log(\eta\bar{\eta})}e^{-\frac{M}{2}\log(\eta/\bar{\eta})} e^{J_{3} \log(\alpha/\eta)}e^{\tilde{J}_{3} \log(\bar{\alpha}/\bar{\eta})}.
\end{equation}
This is the desired weight for the four point functions. Note that this derivation is based on symmetry alone and does not depend on the correctness or not of the skeleton prescription. It assumes a similar form as in $\mathcal{N}=4$ SYM where the R-charge term mixes the sphere and spatial cross-ratios.

Given all of this one could try to compute the first correction at order $\mathcal{O}(h)$ for the four point functions of half-BPS operators. This happens when we insert a single massive mirror magnon. In the course of the computation three problems arise. The first problem occurs when computing the integrand for the mirror correction. As argued before the sum over flavours turn into a transfer matrix. However for four point functions this transfer matrix is twisted due to the insertion of $\mathcal{W}_{\psi}$ weights. Indeed consider the insertion of a L mirror magnon in the edge $1-4$ of length $l$, its contribution is
\begin{align}
    \sum_{Q=1}^{\infty} \int_{-\infty}^{\infty} \frac{du}{2\pi} \mu_Q (u^\gamma) e^{-\tilde{E}_{Q} l} e^{-\frac{E_Q}{2}\log(\eta\bar{\eta})}e^{-\frac{Q}{2}\log(\eta/\bar{\eta})} \times \nonumber \\ \times \textrm{Tr}_{Q}\left( (-1)^F e^{2J_{3} \log(\alpha/\eta)}e^{2\tilde{J}_{3} \log(\bar{\alpha}/\bar{\eta})} \right).
\end{align}
Where we have both the physical ($E_Q$) and mirror ($\tilde{E}_Q$) energies in this expression, $(-1)^{F}$ is the fermion number operator, and the trace is taken over all states in the bound state supermultiplet of mass $Q$. Note that here we are considering the four point function without any excitations, that is why we have this simple trace. 

How to compute this transfer matrix is more an art than science. If one consider the excitations in the string frame we have that they do not carry R-charge, because due to the lightcone gauge fixing the string length (or R-charge) is fixed. Then this trace vanishes due to supersymmetry, which is undesired. However working in the spin chain frame one has that the excitations carry R-charge and in fact we have a dynamical spin chain that changes length when acting with the supercharges. To deal with these length changing effects in the states in \cite{Beisert:2006qh} it was introduced the so called $\mathcal{Z}$-markers that delete or add units of R-charge. Now for the computation of $n$-point correlation functions in $AdS_5 \times S^5$ from hexagonalization it appears that the same $\mathcal{Z}$-markers are necessary. In \cite{Fleury:2016ykk} the authors showed that without assigning $\mathcal{Z}$-markers to mirror states one could not reproduce the 1-loop box integral that appeared from perturbation theory. Turns out that this assignment is \textit{ad hoc} and there is no reason, up to now, for why it works \cite{Fleury:2017eph,Fleury:2020ykw}. Since we do not have perturbative data such a procedure is not possible for us. In summary the hexagon formalism is some sort of hybrid between the string and spin chain frames and understanding it better in the $\mathcal{N}=4$ SYM would allows us to analyze its effect on our setup.

The second problem is the singularity structure of the correlator. Let's for the moment ignore the $\mathcal{Z}$-marker \textit{conundrum}. For one mirror particle the $\mathcal{O}(h)$ correction for L-particles can be computed using the weak coupling expansion of App. \ref{app:weak}. Then we have that the integral part is simply
\begin{equation}
    -\int_{-\infty}^{\infty} \frac{du}{2\pi} \frac{h}{2\left( u^2 + Q^2 /4\right)}   e^{-iu\log(\eta\bar{\eta})},
\end{equation}
which can be computed by residue and yields after the sum over bound states
\begin{equation}
    \frac{h}{2} \log\left(1-\frac{1}{\bar{\eta}}\right).
\end{equation}
This can be found assuming that whatever the $\mathcal{Z}$-marker prescription gives it is $Q$-independent. For R-particles it is the same but with conjugated variables. Since we are in a 2D theory it is expected that the correlators possess singularities like $\log(\eta\bar{\eta})$ in perturbation theory due to the $(\eta\bar{\eta})^{\Delta}$ term in the 2D conformal block expansion \cite{Poland:2018epd}. This could be true here only after one joins the L and R contributions, however there is the problem of the undetermined $\mathcal{Z}$-marker prescription when doing this. Nonetheless it is consistent that the first mirror correction is of order $\mathcal{O}(h)$ since this would correspond to the massless modes' anomalous dimensions that are of the same order.

The third problem are the massless modes. One can see that unless the $\mathcal{Z}$-marker prescription for the two massless multiplets differs all the massless mirror corrections would vanish. Just like in the structure constant case. This would be desirable since otherwise we would have an infinite number of mirror corrections that enters at order $\mathcal{O}(h^0)$. But at the moment there is no way to settle which option would be correct.

\section{Conclusion}
\label{sec:conclusion}

In this work we analyzed in more detail the hexagonalization of the pure RR limit of $AdS_3 \times S^3 \times T^4$ duality. Given the framework introduced in \cite{Eden:2021xhe} we computed mirror corrections and showed how they cancel for the chiral ring structure constants that are protected. We analyzed the coupling dependence of the mirror corrections and how to define them in terms of $\mathfrak{psu}(1|1)_{c.e.}^{\oplus 2}$ transfer matrices. This complements the program initiated in \cite{Eden:2021xhe} to compute structure constants by hexagonalization. Also we gave the first step into computing four point functions in the dual CFT$_2$. Which at the moment there is no alternative way to compute them. In this context we defined the weights $\mathcal{W}_{\psi}$ and showed what are the problems faced in the computation that should be overcomed.

Integrability in the $AdS_3/$CFT$_2$ duality is less explored than the $AdS_5 \times S^5$ case and due to this there are many directions to follow. One of them that is directly related to this work is the pure NSNS hexagonalization. In \cite{Eden:2021xhe} the hexagon framework was introduced only for mixed or pure RR fluxes. The extension of our computations to mixed flux is direct. It amounts to redefine the Zhukovsky variables and slightly change crossing rules. However in these backgrounds the dressing phases of the S-matrices are unknown and thus the hexagon framework is less understood. That is why we stick with the pure RR limit for now. For the pure NSNS regime the hexagons can not be defined. The reason for this is that the central charge of the symmetry algebra of the hexagon vanishes. Then even the two particle hexagon is not fixed by symmetry. The same happens to the pure NSNS S-matrix, which was guessed and then checked to be correct. Nonetheless the spectral problem and CFT$_2$ dual are much simpler in this background and there are a lot of data to compare. This is a direction we are exploring now.

In the four point computation we presented three problems that appeared: ambiguity of the $\mathcal{Z}$-marker prescription, the analytic structure, and the role of the massless modes. One possible origin for these problems is that the initial assumption that the skeleton graphs are given just by Wick contractions may be wrong. One evidence for this is that for the Sym$_N (T^4)$ the correlators are given by covering surfaces and not Wick contractions \cite{Pakman:2007hn,Pakman:2009mi}. How to change this prescription to accomodate these facts is not clear. One first should understand how the pure RR CFT$_2$ is a deformation of Sym$_N (T^4)$. This is another direction that we are exploring. After this is done a way to check if the result is correct is to do the OPE decomposition of the four point correlator and then extract the CFT data as it was done for $\mathcal{N}=4$ SYM in \cite{Coronado:2018ypq}. Then one compares the result with the data from ABA and hexagonalization.

Integrability is very powerful but not \textit{all-powerful}. In the context of $AdS/$CFT its use is a mix of well established techniques from other contexts and a lot of physical intuition to make the correct prescriptions. Due to this it must always be tested against data to be corrected and enhanced. One of the main problems of $AdS_3 \times S^3 \times T^4$ duality is the lack of perturbative data for the dual CFT$_2$ at generic fluxes. One should better understand it in order to amend the problems we detailed here and make progress in these matters.  

\begin{acknowledgments}

We would like to thank Pedro Vieira, Carlos Bercini, Alexandre Homrich, Alessandro Sfondrini, Enrico Olivucci, Francesco Aprile, Leonardo Rastelli, Shlomo Razamat, Tiago Fleury, Frank Coronado, and Vasco Gonçalves for illuminating discussions. This work was supported by FAPESP grant 2019/12167-3 and ICTP-SAIFR FAPESP grant 2016/01343-7. We would also like to thank the organization of Bootstrap 2022 at Porto where this work was finalized for the amazing support.

\end{acknowledgments}

\appendix

\section{Crossing and Mirror Transformations}
\label{app:mirror}

In this section we discuss mirror and crossing transformations in this model. Due to the inclusion of massless modes this slightly differs from the $AdS_5/$CFT$_4$ case. This discussion follows the TBA recently analyzed in \cite{Frolov:2021bwp}. Let's consider first the Zhukovsky variables defined by:
\begin{equation}
    x(u) = \frac{u+i\sqrt{4-u^2}}{2}.
    \label{eq: zhukovsky}
\end{equation}
Then we have the shifted variables:
\begin{equation}
    x^{[\pm a]} (u) = x\left( u \pm\frac{ i a}{h} \right).
\end{equation}
For fundamental massive particles one uses $a=1$ and for massless $a\rightarrow 0$. Here one chooses the branches of the Zhukovsky variables to have $|x^{\pm} (u)|\geq 1$ and $|x(u)|=1$ for physical particles. This restricts $u\in\mathbb{R}$ for massive states and $u\in [-2,2]$ for massless states. 

This parametrization is a good choice since at weak coupling it yields
\begin{equation}
    x^{\pm} (u) = \frac{u\pm i/2}{h} - \frac{h}{u\pm i/2} - \frac{h^3}{\left(u\pm i/2 \right)^3} + \mathcal{O}(h^5).
\end{equation}
Which is the same weak coupling expansion as in \cite{Baggio:2017kza} for the spin chain analysis of the model. To obtain this expansion we have to redefine the spectral parameter $u\rightarrow 2u/h$ and the coupling $h\rightarrow 2h$. For the massless particles we chose $x(u)$ with no coupling dependence or rescaling, it is just the limit $a\rightarrow 0$ of eq. (\ref{eq: zhukovsky}). Another possible choice of parametrization for massless particles is $x$ directly, which would mean that the physical region is the unit circle in the complex plane.

\begin{figure}
\includegraphics[width=0.45\textwidth]{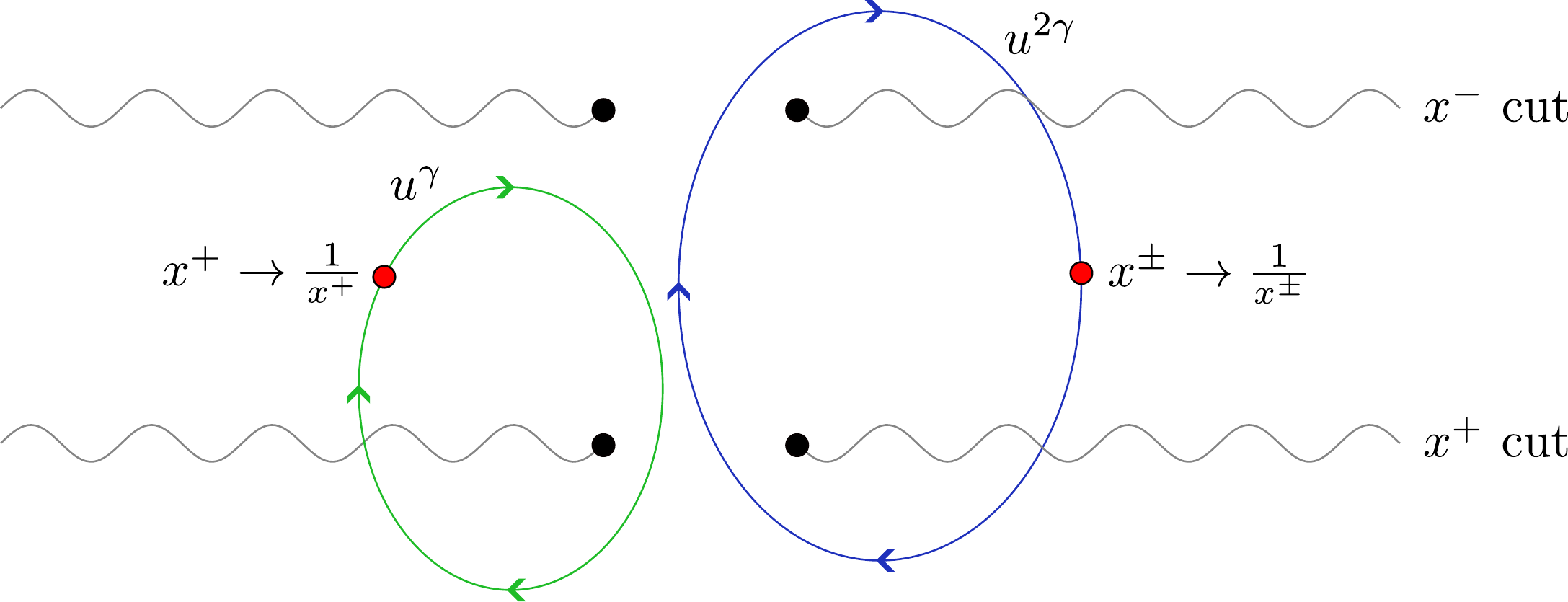}
\caption{Analytical structure of the Zhukovsky variables in the $u$-plane. Here we have the $x^{\pm}$ branch cuts that go out to infinity. In the right (blue) we have a path that denotes the analytic continuation that corresponds to crossing. In the left (green) we have a path corresponding to a mirror transformation. Note that the branch cuts are separated by $i$ and that in the massless limit of the Zhukovsky variables they collide and yield a single branch cut.}
\label{fig:analytical-structure}
\end{figure}

Let's consider the analytical structure in the $u$-plane induced by this parametrization. Note that as usual $x(u)$ has two long cuts starting at $u=\pm 2$. When one crosses this cut $x(u)\rightarrow 1/x(u)$. The analytical structure is shown in Fig. \ref{fig:analytical-structure} \footnote{In \cite{Frolov:2021fmj} the authors introduced a slightly different parametrization of the Zhukovsky variables which leads to different positions for the cuts in the physical and mirror regions, however this is just a matter of convention and one can choose any of the two options to work with.}. The energy and momentum for massive modes are given by
\begin{align}
    & E(u) = \frac{h}{2i}\left( x^{+} - \frac{1}{x^{+}} - x^{-} + \frac{1}{x^{-}} \right), \\ 
    & p(u) = \frac{1}{i} \log\left( \frac{x^{+}}{x^{-}} \right).
\end{align}
To obtain the energy and momentum for bound states we just do $x^{\pm}\rightarrow x^{[\pm Q]}$. In the physical region is easy to see that energy is positive and momentum real and  periodic $-\pi<p<\pi$. So given the analytical structure we note that both quantities have two sets of long cuts, a lower one induced by $x^{+}$ and an upper one by $x^{-}$. As usual when we cross both cuts we do a crossing transformation where $E\rightarrow -E$ and $p\rightarrow -p$. So crossing for massive particles is implemented.

The same analysis can be done for massless particles. We denote their Zhukovsky variables by
\begin{equation}
    x^{+}_{\circ}(u) = x(u) \ \ \ \textrm{and} \ \ \ x^{-}_{\circ}(u) = \frac{1}{x(u)}.
\end{equation}
So they satisfy the constraint $ x^{+}_{\circ}(u)  x^{-}_{\circ}(u) = 1$ and then all functions of the massless Zhukovsky variables can be expressed only in terms of $ x(u)$. Thus the energy and momentum are simply the $a\rightarrow 0$ limit of the massive ones:
\begin{equation}
   E(u) = \frac{h}{i}\left( x - \frac{1}{x} \right) \ \ \ \textrm{and} \ \ \  p(u) = \frac{1}{i} \log\left(x^2 \right).
\end{equation}
In the physical region, energy is non-negative and momentum is periodic as expected. For the analytic structure we note that in the massless limit the cuts on $x^{[+a]}$ and $x^{[-a]}$  combine and there is only two long cuts at $|u|\geq 2$. When these are crossed we do a crossing transformation. Therefore the full rule for crossing transformation is:
\begin{equation*}
    \textrm{\textbf{Crossing:}} \  \ \ x^{+}\rightarrow \frac{1}{x^{+}}, \ \ \ x^{-}\rightarrow \frac{1}{x^{-}}, \ \ \ x\rightarrow \frac{1}{x}.
\end{equation*}
Up to now we have the same as in $AdS_5/$CFT$_4$ case.

Now the mirror kinematics. This is a transformation such that $E\rightarrow i\tilde{p}$ and $p\rightarrow i\tilde{E}$ where $\tilde{E}$ and $\tilde{p}$ are the mirror energy and mirror momentum, respectively. The mirror energy must be positive, but the mirror momentum is real and non periodic. For massive particles the mirror kinematics can be found by crossing only the lower $x^{+}$ cut. Therefore
\begin{align}
    & \tilde{E}(u) = \log\left( x^{+} x^{-} \right), \\
    & \tilde{p}(u) = -\frac{h}{2}\left( \frac{1}{x^{+}} - x^{+} - x^{-} + \frac{1}{x^{-}} \right).
\end{align}
For $u\in\mathbb{R}$ these satisfy the physical requirements. Then mirror kinematics here works just like in $AdS_5$ case. 

However for massless particles since both cuts merge it  does not appear that there is a \textit{half} crossing transformation. Nonetheless one can analytic continue $E$ and $p$ to another region where the kinematics is the one for the mirror theory. Indeed given
\begin{align}
     & \tilde{E}(u) = -\log\left(x^2 \right), \\ 
     & \tilde{p}(u) = -h\left( x - \frac{1}{x} \right),
\end{align}
the region where these satisfy the physical requirements is $|u|\geq 2$. But with the care that we are below the cuts, since this is the region where $\tilde{E}$ is positive. So the mirror transformation here is:
\begin{eqnarray}
    \textrm{\textbf{Mirror:}} 
    \  \ \ x^{+}\rightarrow \frac{1}{x^{+}} \ \textrm{for}\  M\geq 1, \nonumber \\
    u\in [-2,2] \rightarrow u+i0^{-},\ |u|\geq 2 \ \textrm{for} \ M=0. \nonumber
\end{eqnarray}
This is the mirror kinematics we used for the wrapping corrections.

\section{Bound States in $AdS_3/$CFT$_2$}
\label{app:bs}

In this appendix we describe the bound states in this background and how the fusion works. A bound state of $n$ particles is formed when a composite state has energy and momentum of each constituent arranged in a way that it transforms in a single particle representation of $\mathfrak{psu}(1|1)^{\oplus 4}_{c.e.}$. As detailed in \cite{Frolov:2021bwp,Borsato:2013hoa} the bound states can be composed only of L or R particles. Two particle representations build of L-R or massless-massless or L/R-massless excitations do not make an appropriate one particle representation of the superalgebra for any set of momenta. A nice fact of $AdS_3/$CFT$_2$ is that the bound states in this model transform in the same representation of the fundamental particles, unlike the $AdS_5/$CFT$_4$ case \cite{Frolov:2021bwp}.

Given $n$ fundamental particles we can built a bound state by fusion. Consider the following set of rapidities
\begin{equation}
    u^{(n)}_j = u+ i\frac{(n+1-2j)}{2}, \ \ \textrm{where} \ \ j=1,\cdots,n.
\end{equation}
Then the tensor product of $n$ fundamental particles where each one has rapidity $u_{j}^{(n)}$ behaves as a single particle. The bound state has $\mathfrak{u}(1)$ charge $M=n$ or $M=-n$ for L and R particles, respectively. By summing the energies and momentum of the constituents we obtain $E_{Q}(u)$ and $p_{Q} (u)$. 

Instead of energy and momentum we can also fuse the scalar factors of the hexagon following the procedure of \cite{Fleury:2016ykk}. Unlike the $AdS_5$ case, now we have multiple hexagon scalar factors $h^{LL}$, $h^{RR}$, $h^{LR}$, $h^{RL}$, $h^{* \circ}$, $h^{\circ *}$, $h^{\circ\circ}$ where $*=L$ or $*=R$. This happens because now we have four supermultiplets interacting. Thus let's first establish some properties of these scalar factors. From L$/$R-symmetry we have the identities:
\begin{equation}
    h^{LL} = h^{RR} \ \ \textrm{and} \ \ h^{RL} = \sqrt{\frac{x^{-}_{p} x^{-}_{q}}{x^{+}_{p} x^{+}_{q}}} \frac{1-x^{+}_{p} x^{+}_{q}}{1- x^{-}_{p} x^{-}_{q}} \ h^{LR}.
\end{equation}
Also we have that
\begin{align}
    h^{L \circ} (p,q) h^{\circ L} (q,p) = & h^{R \circ} (p,q) h^{\circ R} (q,p) =  \nonumber \\ 
    &  = \frac{x^{+}_{p}-x_{q}}{x^{-}_{p}-x_{q}} \frac{1-x^{-}_{p} x_{q}}{1-x^{+}_{p} x_{q}}.
\end{align}
Then we define four classes of independent scalar factors
\begin{equation}
    h^{\bullet\bullet} \equiv h^{LL}, \ \ \Tilde{h}^{\bullet\bullet} \equiv h^{LR}, \ \ h^{\bullet\circ}_{*} \equiv h^{*\circ}, \ \ h^{\circ\bullet}_{*} \equiv h^{\circ *}.
\end{equation}
Where $* = L,R$. Note that here we do not have $h^{\bullet\circ}_{L} = h^{\bullet\circ}_{R} $ or $h^{\circ\bullet}_{L} = h^{\circ\bullet}_{R} $. These are genuinely distinct.

The fusion formula for these scalar factors is simple to define and we find:
\begin{equation}
    h^{\bullet\bullet}_{Q_1 Q_2 } (u,v) = \prod\limits_{m=1}^{Q_1} \prod\limits_{n=1}^{Q_2} h^{\bullet\bullet} (u_{m}^{( Q_1 )},v_{n}^{( Q_2 )}).
\end{equation}
\begin{equation}
    \Tilde{h}^{\bullet\bullet}_{Q \Bar{Q} } (u,v) = \prod\limits_{m=1}^{Q} \prod\limits_{n=1}^{\Bar{Q}} \tilde{h}^{\bullet\bullet} (u_{m}^{( Q )},v_{n}^{( \Bar{Q} )}).
\end{equation}
\begin{align}
    h_{Q \Bar{Q} }^{RL}(u,v) = &  \prod\limits_{m=1}^{Q}  \prod\limits_{n=1}^{\Bar{Q}} \biggl( \sqrt{\frac{x^{-} ( u_{m}^{( Q )} ) x^{-} (v_{n}^{( \Bar{Q} )}) }{x^{+} (u_{m}^{( Q )}) x^{+} (v_{n}^{( \Bar{Q} )}) }} \times \nonumber \\ 
    & \times \frac{1-x^{+} (u_{m}^{( Q )}) x^{+}(v_{n}^{( \Bar{Q} )})}{1- x^{-} (u_{m}^{( Q ) }) x^{-}(v_{n}^{( \Bar{Q} )})} \biggr) \Tilde{h}^{\bullet\bullet}_{Q \Bar{Q} } (u,v).
\end{align}
\begin{equation}
    h_{*,Q}^{\bullet\circ} (u,v) = \prod\limits_{m=1}^{Q} h_{*}^{\bullet\circ} (u_{m}^{( Q )},v).
\end{equation}
\begin{equation}
    h_{*,Q}^{\circ\bullet} (u,v) = \prod\limits_{n=1}^{Q} h_{*}^{\circ\bullet} (u,v_{n}^{( Q )}).
\end{equation}
Similar fusion formulas could be found for the dressing factors of the S-matrix. These are useful to compute higher orders in the $h$-expansion for structure constants or correlation functions.

\section{Weak Coupling Expressions}
\label{app:weak}

Here we compile the weak coupling expression of some useful quantities. For the measure:
\begin{align}
    & \mu_Q (u) =
    \frac{1}{2 Q}-\frac{Q}{4\left( u^2 + Q^2 /4\right)^2} h^2 +\frac{ Q \left(3 Q^2-32
   u^2\right)}{16\left( u^2 + Q^2 /4\right)^4}  h^4 \nonumber \\ 
   & -\frac{Q \left(5 Q^4-160 Q^2 u^2+384
   u^4\right)}{32 \left( u^2 + Q^2 /4\right)^6} h^6 +\mathcal{O}\left(h^6\right), \\
   & \mu_Q (u^\gamma) =
    -\frac{h}{2\left( u^2 + Q^2 /4\right)}+\frac{  \left(Q^2-8 u^2\right)}{4\left( u^2 + Q^2 /4\right)^3} h^3 \nonumber \\
   & -\frac{
    \left(3 Q^4-80 Q^2 u^2+128 u^4\right)}{16 \left( u^2 + Q^2 /4\right)^5} h^5 + \mathcal{O}\left(h^7\right), \\
    & \mu_\circ (u) = \frac{1}{4-u^2}.
\end{align}

For the energy and momentum in the physical region: 
\begin{align}
    & E_Q (u) = 
    Q+\frac{2  Q}{ u^2 + Q^2 /4} h^2 -\frac{ Q\left(Q^2-12 u^2\right)}{2 \left( u^2 + Q^2 /4\right)^3} h^4 \nonumber \\ 
    & +\frac{Q \left(Q^4-40 Q^2 u^2+80  u^4\right)}{4 \left( u^2 + Q^2 /4\right)^5} h^6 +\mathcal{O}\left(h^7\right), \\
    & e^{ i p_Q (u)} = 
    \frac{u+i Q /2}{u - i Q /2}+\frac{2 i  Q u}{\left( u^2 + Q^2 /4\right) (u-i Q/2)^2} h^2 \nonumber \\ 
    & -\frac{Q u \left(3 i
   Q^2+4 Q u-12 i u^2\right)}{2 \left( u^2 + Q^2 /4\right)^3 (u-i Q/2)^2} h^4 +\mathcal{O}\left(h^5\right), \\
   & E_{\circ} (u) = h \sqrt{4-u^2}, \\
   & e^{ i p_{\circ} (u)} = 
    \frac{u^2 - 2 +i u \sqrt{4-u^2}}{2}.
\end{align}
Now in the mirror kinematics we have:
\begin{align}
    & e^{-\tilde{E}_Q (u)} = 
    \frac{h^2}{u^2 + Q^2 /4} -\frac{  \left(Q^2-4 u^2\right)}{2 \left(u^2 + Q^2 /4\right)^3} h^4 \nonumber \\
    &  +\frac{
   \left(5 Q^4-88 Q^2 u^2+80 u^4\right)}{16 \left(u^2 + Q^2 /4\right)^5}  h^6 + \mathcal{O}\left(h^7\right), \\
   & \tilde{p}_Q (u) = 
    2 u-\frac{4  u}{u^2 + Q^2 /4} h^2 +\frac{\left(3 Q^2 u-4 u^3\right)}{ \left(u^2 + Q^2 /4\right)^3} h^4 \nonumber \\
    & -\frac{\left(5 Q^4 u-40 Q^2 u^3+16 u^5\right)}{2 \left(u^2 + Q^2 /4\right)^5} h^6 +\mathcal{O}\left(h^7\right).
\end{align}
For massless mirror particles we just do the analytic continuation in $u$ and obtain similar expressions to this.

\section{Transfer Matrices in $AdS_3/$CFT$_2$}
\label{sec:transfermatrix}

In this appendix we write the transfer matrix eigenvalues $\hat{\Delta}^{*}$ in terms of the $\mathfrak{psu}(1|1)_{c.e.}^{\oplus 2}$ S-matrix elements given in \cite{Seibold:2022mgg}. These are
\begin{align}
    &\hat{\Delta}^L (u | \{ u_{j} \} ) =  \prod_{j=1}^{K_L} A^{LL}_{u,u_{j}} \prod_{j=1}^{K_R} C^{LR}_{u,u_{j}} \prod_{j=1}^{K_{\circ}} A^{L\circ}_{u,u_{j}} + \nonumber \\
     & -(-1)^{K_R} \prod_{j=1}^{K_L} D^{LL}_{u,u_{j}} \prod_{j=1}^{K_R} E^{LR}_{u,u_{j}} \prod_{j=1}^{K_{\circ}} D^{L\circ}_{u,u_{j}}, \\
     &\hat{\Delta}^R (u | \{ u_{j} \} ) = (-1)^{K_R+1} \prod_{j=1}^{K_L} D^{RL}_{u,u_{j}} \prod_{j=1}^{K_R} F^{RR}_{u,u_{j}} \prod_{j=1}^{K_{\circ}} D^{R\circ}_{u,u_{j}} + \nonumber \\
     & + \prod_{j=1}^{K_L} A^{RL}_{u,u_{j}} \prod_{j=1}^{K_R} B^{RR}_{u,u_{j}} \prod_{j=1}^{K_{\circ}} A^{R\circ}_{u,u_{j}}, \\
     & \hat{\Delta}^{\circ} (u | \{ u_{j} \} ) = 0.
\end{align}
For massless modes the transfer matrix vanishes since the two massless multiplets in $\mathfrak{psu}(1|1)_{c.e.}^2$ are equal but come with opposite statistics for the excitations \cite{Eden:2021xhe}. Also to find the transfer matrices for bound states $\Delta_{a}^{*}$ we just change $x^{\pm}(u)\rightarrow x^{[\pm a]}(u)$ in the S-matrix elements.

\section{Phase Factors For Zero Momentum Massless Fermions Scattering}

\label{sec:phasefactors}

In this appendix we compute the phase factors $S_{\circ A}(p)$. We prove that they are flavour independent for each multiplet and also that the S-matrices are indeed pure transmission. In this section we will use the momentum parametrization of the Zhukovsky variables. For the massless particles this is simply:
\begin{equation}
    x(p) = e^{ip/2}\ \textrm{sgn}(p).
\end{equation}
Then the zero momentum limit is either $x (0^{\pm}) = \pm 1$ depending if we approach from positive or negative momentum.

The massless fermions can be written in terms of $\mathfrak{psu}(1|1)_{c.e.}^2$ excitations. Indeed:
\begin{equation}
    \begin{array}{cc}
        |\chi^1 \rangle = | \phi^{B}_{\circ}  \otimes \phi^{F}_{\circ} \rangle  & \ \ \  |\chi^2 \rangle = | \phi^{F}_{\circ}  \otimes \phi^{B}_{\circ} \rangle \\
        |\tilde{\chi}^1 \rangle = | \varphi^{F}_{\circ}  \otimes \varphi^{B}_{\circ} \rangle  & \ \ \ \  |\tilde{\chi}^2 \rangle = | \varphi^{B}_{\circ}  \otimes \varphi^{F}_{\circ} \rangle,
    \end{array}
\end{equation}
Where $\phi_{\circ}^{*}$ and $\varphi_{\circ}^{*}$ denote highest and lowest states of the $\mathfrak{psu}(1|1)_{c.e.}^{\oplus 2}$ superalgebra, respectively. For more details see \cite{Eden:2021xhe}. Therefore in the hexagon computation involving massless zero momentum fermions one has to include the following $\mathfrak{psu}(1|1)_{c.e.}^2$ excitations:
\begin{equation}
    \label{eq:nullmomenta}
    \phi_{\circ}^{F} (0^+), \ \ \varphi_{\circ}^{B} (0^+), \ \ \phi_{\circ}^{B} (0^-), \ \ \textrm{and} \ \  \varphi_{\circ}^{F} (0^-).
\end{equation}
Where the zero momentum limit for each excitation is chosen in a way that each state have the appropriate chirality.

Let $X^{(L,R)}$ be a particle in a massive L/R-multiplet. Their scattering with the zero momentum insertions can be computed by just doing the appropriate zero momentum limit on each S-matrix element. Using the S-matrices in \cite{Eden:2021xhe} we find:
\begin{align}
    & \mathcal{S}|\phi_{\circ}^{B} X^{(L,R)}_p\rangle = |X^{(L,R)}_p \phi_{\circ}^{B} \rangle, \\
    & \mathcal{S}|\phi_{\circ}^{F} X^{(L,R)}_p\rangle = (-1)^{|X^{(L,R)}|} |X^{(L,R)}_p \phi_{\circ}^{F} \rangle, \\
    & \mathcal{S}|\varphi_{\circ}^{F} X_p\rangle = 
    \begin{cases}
        (-1)^{|X^{(L)}|} e^{ip/2} \frac{1+x^{-}_{p}}{1+x^{+}_{p}} |X^{(L)}_p \varphi_{\circ}^{F} \rangle \\
        (-1)^{|X^{(R)}|} e^{-ip/2} \frac{1+x^{+}_{p}}{1+x^{-}_{p}} |X^{(R)}_p \varphi_{\circ}^{F} \rangle
    \end{cases}, \\
     & \mathcal{S}|\varphi_{\circ}^{B} X_p\rangle = 
    \begin{cases}
        e^{ip/2} \frac{1-x^{-}_{p}}{1-x^{+}_{p}} |X^{(L)}_p \varphi_{\circ}^{B} \rangle \\
        e^{-ip/2} \frac{1-x^{+}_{p}}{1-x^{-}_{p}} |X^{(R)}_p \varphi_{\circ}^{B} \rangle
    \end{cases}.
\end{align}
With $|X| = 0$ or $|X|=1$ if $X$ is a boson or a fermion, respectively. Also the massless particles have the null momenta taken as in eq. (\ref{eq:nullmomenta}). For massless particles $X^{\circ}$ we find:
\begin{align}
    & \mathcal{S}|\phi_{\circ}^{B} X^{\circ}_{p}\rangle = |X^{\circ}_{p} \phi_{\circ}^{B} \rangle, \\
    & \mathcal{S}|\phi_{\circ}^{F} X^{\circ}_{p}\rangle = (-1)^{|X^{\circ}|} |X^{\circ}_{p} \phi_{\circ}^{F} \rangle, \\ 
    & \mathcal{S}|\varphi_{\circ}^{F} X^{\circ}_{p} \rangle = (-1)^{|X^{\circ}|} e^{ip/2} \frac{1+1/x_{p}}{1+x_{p}} |X^{\circ}_{p} \varphi_{\circ}^{F} \rangle, \\
     & \mathcal{S}|\varphi_{\circ}^{B} X^{\circ}_{p} \rangle =  e^{ip/2} \frac{1-1/x_{p}}{1-x_{p}} |X^{\circ}_{p} \varphi_{\circ}^{B} \rangle.
\end{align}
Clearly these S-matrices are pure transmission and flavour independent as expected.


\bibliography{AdS3bib}

\end{document}